\documentclass[conference]{IEEEtran}
\usepackage{cite}
\usepackage{amsmath,amssymb,amsfonts}
\usepackage{algorithmic}
\usepackage{graphicx}
\usepackage{textcomp}
\usepackage{xcolor}
\usepackage{tcolorbox}
\usepackage{booktabs}
\usepackage{multirow}
\usepackage[normalem]{ulem}
\useunder{\uline}{\ul}{}
\def\BibTeX{{\rm B\kern-.05em{\sc i\kern-.025em b}\kern-.08em
    T\kern-.1667em\lower.7ex\hbox{E}\kern-.125emX}}
\usepackage{environ} 
 
\NewEnviron{NORMAL}{%
    \scalebox{1.5}{$\BODY$} 
} 

\begin{document}

\title{Ollabench: Evaluating LLMs' Reasoning for Human-centric Interdependent Cybersecurity*\\
{\footnotesize \textsuperscript{*} The final conference/journal version may have significantly more content updates.}
\thanks{This paper does not reflect the views of the US FDA nor the US Government.}
}

\author{\IEEEauthorblockN{Tam n. Nguyen}
\IEEEauthorblockA{\textit{tom.nguyen@ieee.org} \\
https://orcid.org/0000-0002-8577-8342}
}

\maketitle

\begin{abstract}
Large Language Models (LLMs) have the potential to enhance Agent-Based Modeling by better representing complex interdependent cybersecurity systems, improving cybersecurity threat modeling and risk management. However, evaluating LLMs in this context is crucial for legal compliance and effective application development. Existing LLM evaluation frameworks often overlook the human factor and cognitive computing capabilities essential for interdependent cybersecurity. To address this gap, I propose OllaBench, a novel evaluation framework that assesses LLMs' accuracy, wastefulness, and consistency in answering scenario-based information security compliance and non-compliance questions. OllaBench is built on a foundation of 24 cognitive behavioral theories and empirical evidence from 38 peer-reviewed papers. OllaBench was used to evaluate 21 LLMs, including both open-weight and commercial models from OpenAI, Anthropic, Google, Microsoft, Meta and so on. The results reveal that while commercial LLMs have the highest overall accuracy scores, there is significant room for improvement. Smaller low-resolution open-weight LLMs are not far behind in performance, and there are significant differences in token efficiency and consistency among the evaluated models. OllaBench provides a user-friendly interface and supports a wide range of LLM platforms, making it a valuable tool for researchers and solution developers in the field of human-centric interdependent cybersecurity and beyond.
\end{abstract}


\section{Introduction} \label{sec_introduction}
Most CEOs care about maintaining the right level of cybersecurity at an optimal cost. With the cyber threat landscape keeps evolving, organizations cannot manage cybersecurity risks effectively despite increased cybersecurity investments  \cite{McLennan2021TheEdition}. This challenge can be explained via Interdependent Cybersecurity (IC)\cite{Kunreuther2003InterdependentSecurity}. For the first reason, optimizing cybersecurity investments in existing large interdependent systems is a well-known non-convex hard problem with no effective solution \cite{Mai2021OptimalDesign}. Second, smaller systems are growing in complexity and interdependence \cite{Brumfield2021WhyDangerous.}. Last, new low frequency, near simultaneous, macro-scale risks such as global pandemics, financial shocks, and geopolitical conflicts have compound effects on IC \cite{bremen-2022}.

\begin{figure*}[htbp]
\centerline{\includegraphics[width=\textwidth,height=\textheight,keepaspectratio]{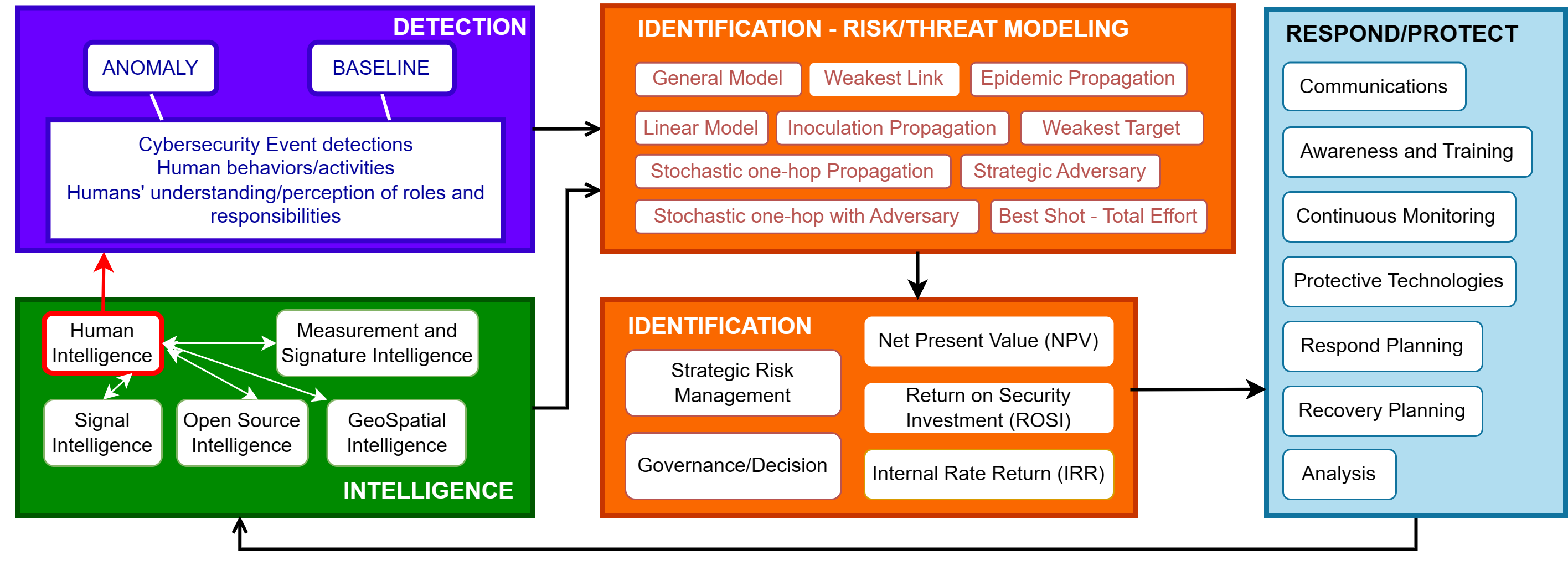}}
\caption{The Human-centered Application Areas for Interdependent Cybersecurity}
\label{fig_HIC_Framework}
\end{figure*}

One way to increase effectiveness in IC risk management is hardening the key components shared by all large to small systems. With humans account for half of the problems in IC \cite{Kianpour2021SystematicallySurvey, Laszka2014AGames}, hardening the human factors is an obvious strategy. Figure \ref{fig_HIC_Framework} summarized the US NIST Cybersecurity Framework \cite{Barrett2018Framework1.1}, Kianpour et al. \cite{Kianpour2021SystematicallySurvey}, Andrade et al. \cite{Andrade2019CognitiveCybersecurity}, and Laszka et al. \cite{Laszka2014AGames} into main application areas of Human-centric IC. These application areas ultimately support the common goal of staying at least one step ahead of cybersecurity threats.

Specifically, the "Risk and Threat Modeling" area plays out the interdependent games which involve normal and adversarial players \cite{Laszka2014AGames}. The total player number is $1 \le i \le n$. Cybersecurity investments vector $X = [x_1,...,x_n]^T$ where $x-i$ is the cybersecurity investments of all players except player $x_i$. The risk is $f_i(x) = f_i(x_i,x_{-i}) , f_i \in [0,1]$. With $C_i$ as the unit cost of cybersecurity investment for player $i$, the loss function is $L_i$ and the expected cost is $L_if_i(x) + C_ix_i$. Being ahead of adversaries lowers expected costs by mechanisms such as insurance, rewards, penalties, regulations, audits, coordination, and sharing information, alongside enhancing cybersecurity technology. Cybersecurity Independence Game models’ outputs can be the inputs of Risk Management models, Decision Theory models, and other models. The results of these model will then inform the specific actions which an organization should act on. Finally, the organization must gather the intelligence to measure the effectiveness of respond/protect action plans as well as potential threats. The cycle completes and a new cycle begins.

\begin{table*}[ht]
\caption{Recent Works on Evaluation of LLMs Applicable Towards Interdependent Cybersecurity}
\resizebox{\textwidth}{!}{%
    \begin{tabular}{@{}lrllllll@{}}
    \toprule
    \multicolumn{1}{c}{\textbf{Work}} & \multicolumn{1}{c}{\textbf{Year}} & \multicolumn{1}{c}{\textbf{H}} & \multicolumn{1}{c}{\textbf{C}} & \multicolumn{1}{c}{\textbf{R}} & \multicolumn{1}{c}{\textbf{Type}} & \multicolumn{1}{c}{\textbf{Grounding}} & \multicolumn{1}{c}{\textbf{Novel Contribution}} \\ \midrule
    \textbf{OLLABENCH} & \multicolumn{1}{l}{2024} & $\bullet$ & $\bullet$ & $\bullet$ & Scenarios & 24 Cognitive Behavioral Theories & Knowledge graph, dataset (10000 items), dataset generator, GUI interface \\
    Agrawal et al. \cite{Agrawal2024CyberQ:LLMs} & 2024 &  & $\bullet$ &  & QA & AISecKG & Data set generator (CyberGen) and benchmark dataset (CyberQ - 4000 items) \\
    Bhat et al. \cite{Bhatt2024CyberSecEvalModels} & 2024 &  & $\bullet$ & $\bullet$ & Test case & Mitre ATT\&CK & False refusal rate and propensity of complying with malicious instructions \\
    Garza Assessing & 2023 &  & $\bullet$ &  & QA & Mitre ATT\&CK & Prompt engineering insights in threat behavior domain \\
    Guha et al. \cite{Guha2024Legalbench:Models} & 2024 &  &  & $\bullet$ & Scenarios & Expert knowledge and experience & Benchmark dataset for evaluating LLMs' legal reasoning \\
    Hendrycks et al. \cite{Hendrycks2021ALIGNINGVALUES} & 2021 & $\bullet$ &  &  & Scenarios & Five core ethics theories & The ETHICS benchmark dataset (130,000 items) \\
    Jin et al. \cite{Jin2024Cladder:Models}& 2024 &  &  & $\bullet$ & Scenarios & Existing causal graphs and queries & Benchmark dataset for evaluating LLMs' formal causal reasoning \\
    Jin et al. \cite{Jin2024Crimson:Models} & 2024 &  & $\bullet$ & $\bullet$ & Scenarios & Mitre ATT\&CK & Retrieval-Aware Training with Reason method, Abstract Syntax Trees metric \\
    Kosinki et al. \cite{Kosinski2023EvaluatingTasks}& 2024 & $\bullet$ &  & $\bullet$ & Scenarios & Theory of Mind & Benchmark dataset to measure LLMs' performance on false-belief tasks \\
    Li et al. \cite{Li2024SALAD-Bench:Models}& 2024 &  & $\bullet$ &  & QA \& Scenarios & Existing taxonomies and policies & Multi-level benchmark dataset with attack/defense enhanced scenarios \\
    Scherrer et al. \cite{Scherrer2023EvaluatingLLMs}& 2024 & $\bullet$ &  &  & Scenarios & Five papers on moral psychology & A statistical method for eliciting encoded moral beliefs in LLMs \\
    Sun et al. \cite{Sun2024LLM4Vuln:Reasoning}& 2024 &  & $\bullet$ & $\bullet$ & Scenarios & 75 Smart Contract vulnerabilities & 4950 scenarios to measure LLMs vulnerability reasoning \\
    Yuan et al. \cite{Yuan2024S-Eval:Models}& 2024 &  & $\bullet$ &  & Test case & Self-curated Risk Taxonomy & Expert safety-critique LLM to generate prompt-based evaluation tests \\
    \hline
    \addlinespace 
    \multicolumn{8}{l}{Note: H, C, R indicate the papers' focus on Human Factor, Cybersecurity, and Reasoning respectively.}
    \end{tabular}
    }
\label{tab:lit-review}
\end{table*}

In contrast to mathematical models used in IC interdependent games, Agent-Based Modeling (ABM) is better in capturing individual characteristics \cite{Cuevas2020AnFacilities}. Rooted in methodological individualism, ABM incorporates three principal elements: agents, environment, and rules of interaction \cite{Axtell2022Agent-BasedFuture}. ABM agents represent subsystems and their entities, governed by flexible rules that facilitate the simulation of complex relationships and interactions \cite{An2020Editorial:Models, Silva2020SimulatingReview}. This bottom-up approach allows the micro-level agent behaviors to collectively generate emergent macro-level structures. Therefore, ABM enables the exploration of various complex scenarios at the macro scale, invaluable in assessing the potential development of cybersecurity threats within different large and complex IC contexts.

Large Language Models (LLMs) can be highly effective in developing ABM agents that mimic human networks, human cognitive patterns and articulate decisions in ways comprehensible to humans. As an illustrative example, Park et al. \cite{Park2023GenerativeBehavior} showed a network of 25 LLM-powered agents which can replicate the behavioral economics observed in a human group, complete with discernible cognitive traits. More future LLM-based simulation can be expected. However, in accordance with President Joe Biden's Executive Order 14110 \cite{Biden2023ExecutiveIntelligence}, it is imperative to establish rigorous evaluation mechanisms for the LLMs to ensure they are safe, secure, and trustworthy. The order mandates enhanced scrutiny of AI systems to protect public welfare, prevent discrimination, and ensure privacy and civil liberties are upheld. Table \ref{tab:lit-review} shows recent works on the evaluation of LLMs that are relevant to IC.

We notice that most recent LLM evaluation works for cybersecurity does not involve human factor while LLM reasoning evaluation works do not involve both cybersecurity and human factor. This phenomenon is consistent with the fact that research and applications focusing on human-centered IC remains limited \cite{Kianpour2021SystematicallySurvey} while contemporary LLM evaluation frameworks, such as Stanford's HELM \cite{Bommasani2023HolisticModels}, fail to assess the cognitive computing capabilities of LLMs. For example, HELM includes a broad array of metrics for evaluating various aspects of NLP. However, none of these metrics directly evaluates cognitive computing functions. This oversight may stem from the relative novelty of LLMs, skewed public interest, nascent use cases, and a general lack of understanding of how LLMs can be leveraged in IC risk and threat modeling.

Therefore, OllaBench was born to help both researchers and application developers conveniently evaluate their LLM models within the context of cybersecurity compliance or non-compliance behaviors. In other words, OllaBench helps with answering the question 

\begin{itemize}
    \item \textbf{What is the best LLM for reasoning about human cognitive behavioral observation within the context of cybersecurity compliance or non-compliance?}
\end{itemize}

\begin{figure*}[htbp]
\centerline{\includegraphics[width=\textwidth,height=\textheight,keepaspectratio]{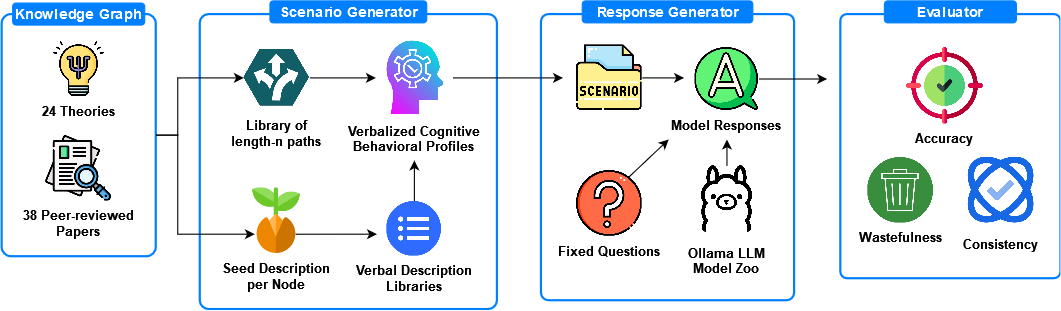}}
\caption{The Design of OllaBench}
\label{fig_design}
\end{figure*}

Figure \ref{fig_design} describes the overall design of OllaBench. Specifically, OllaBench can rank LLMs based on their accuracy, wastefulness, and consistency in answering scenario-based information security compliance/non-compliance questions. Early stage solution developers may leverage OllaBench to identify the best LLM for their human-centric IC application development. Researchers may leverage OllaBench to measure their fine-tuned LLMs' performances.

OllaBench's contributions are many fold.
\begin{itemize}
    \item A novel knowledge graph (Cybonto-4-IC) grounded on 24 cognitive behavioral theories and 38 peer-reviewed papers.
    \item A scenario-based default benchmark dataset with 10,000 items about cognitive behavioral cybersecurity compliance or non-compliance.
    \item A dataset generator powered by Cybonto-4-IC and generative AI for the purposes of robust continuous evaluation and fine-tuning.
    \item An evaluator that support evaluation of all open-weight LLMs as well as commercial LLMs from OpenAI, Google, and Anthropic.
    \item A friendly yet professional Graphical User Interface to support a broad base of OllaBench users.
\end{itemize}

The rest of the paper is organized as follows. Section \ref{sec_background} describes the foundational theories for the paper's evaluation methodology and artifact design. Section \ref{sec_ollabench} presents major aspects of the novel OllaBench. Section \ref{sec_results} describes the results of applying OllaBench in evaluation of 21 LLMs towards information security policy compliant/non-compliance cognitive behavioral reasoning. Section discusses additional important topics before the final conclusion.

\section{Theoretical Background} \label{sec_background}
\subsection{On Evaluation Science}
According to the classical definition, science represents the intellectual and practical endeavor characterized by the methodical examination of the physical and natural realms, achieved through empirical observation and experimental investigation\cite{Patton2018EvaluationScience}. Evaluation pertains to the process of rendering judgments concerning the merit, value, significance, credibility, and utility of various subjects under scrutiny \cite{Patton2018EvaluationScience}. Notably, evaluation Science is confronted with challenges stemming from its interdisciplinary nature and expansive scope which lead to a lack of uniformity in defining what constitutes evaluation \cite{Patton2018EvaluationScience}.

In such context, evaluative criteria are essential for delineating what constitutes a "high quality" or "successful" evaluand. Domain and source are the main evaluative criteria components. The domain refers to the focus or substance of a criterion, encapsulating the specific areas of assessment relevant to the evaluand \cite{Teasdale2021EvaluativeSources}. Conversely, the source denotes the origin of the criterion, whether it be an individual, a collective body, or a documented standard \cite{Teasdale2021EvaluativeSources}.

There are nine domains: Relevance, which assesses the pertinence of the evaluand to its intended objectives; Design, which evaluates the evaluand based on established principles, best practices, standards, and laws; Alignment, which measures the congruence of the evaluand with overarching initiatives; Effectiveness, which gauges the degree to which the evaluand achieves its intended outcomes; Unintended Effects, which considers both positive and negative consequences not originally anticipated; Consequence, which examines the broader impacts of the evaluand; Equity, which assesses the fairness and inclusivity of the evaluand; Resource Use, which evaluates the efficiency and appropriateness of resource allocation; and Sustainability, which considers the long-term viability and environmental impact of the evaluand \cite{Teasdale2021EvaluativeSources}.

Evaluation methodologies in Design Science Research include formative and summative approaches with two execution strategies of artificial, involving laboratory experiments and simulations, and naturalistic, which examines technology performance in real-world settings \cite{Venable2016FEDS:Research}. Formative evaluations aim to generate empirically grounded interpretations to enhance the evaluand's characteristics or performance, primarily focusing on outcomes to facilitate improvement actions \cite{Venable2016FEDS:Research}. Conversely, summative evaluations seek to establish empirically grounded common understandings of the evaluand across varying contexts, concentrating on meanings to inform influential decisions \cite{Venable2016FEDS:Research}. While formative evaluations guide improvements during the development process, summative evaluations assess the results post-development. The utility of formative evaluations lies in their capacity to refine process outcomes, whereas summative evaluations ascertain how well these outcomes align with preset expectations.

Notably, theory-driven evaluation has been increasingly advocated by scholars, practitioners, and organizations, positioning it as a favored methodology in evaluation practices \cite{Coryn2011A2009}. Such theory-driven evaluations are grounded in a clear theory or model that delineates how the program leads to desired or observed outcomes, with the evaluation itself being informed by this model. The theory-driven evaluation's main principles \cite{Coryn2011A2009} are:
\begin{itemize}
    \item A coherent and plausible program theory must be developed.
    \item Evaluations should center and be prioritized following the framework of the program's theory.
    \item The framework should direct the evaluation's planning, design, and implementation, taking into account pertinent contingencies.
    \item Framework constructs must be assessed
    \item Any shortcomings or unintended consequences, effectiveness, and causal links between the theoretical constructs should be identified.
\end{itemize}

\subsection{On Human-centered Metrics}
ISO standards’ definition of human-centredness is anchored in four quality objectives: usability, accessibility, user experience, and the mitigation of use-related risks \cite{Sankowski2023TheActivities}. It underscores the context of use as a preliminary step to designing human-centeredness metric requirements. Additionally, evaluation metric must be versatile enough to serve both researchers focusing on human-centered design principles and product designers.

Generally, metrics are classified into three categories: quantitative-subjective, qualitative, and integrated, with the choice often dictated by the availability and precision of the underlying data\cite{PSU2013MetricsEvaluation}. Data accessibility and affordability are crucial; without these, reliance on assumptions or proxy data may compromise the assessment's validity and reliability. The selection of an appropriate measurement metric is contingent upon the user's specific needs and objectives, the field of study, and the data at hand \cite{PSU2013MetricsEvaluation}. Central to this process is establishing a clear evaluation purpose; absent from this, metrics amount to mere data, lacking actionable values. Additionally, various agencies, companies, and researchers have developed metrics to tackle diverse challenges, often within disparate frameworks \cite{PSU2013MetricsEvaluation}. This has led to a proliferation of assessment methods characterized by differing rationale, terminology, and methodologies, resulting in contrasting outcomes. The design of universally comprehensive metrics, serving as a panacea for decision-makers, proves to be a formidable task.

\begin{figure*}[htbp]
\centerline{\includegraphics[width=370pt,height=\textheight,keepaspectratio]{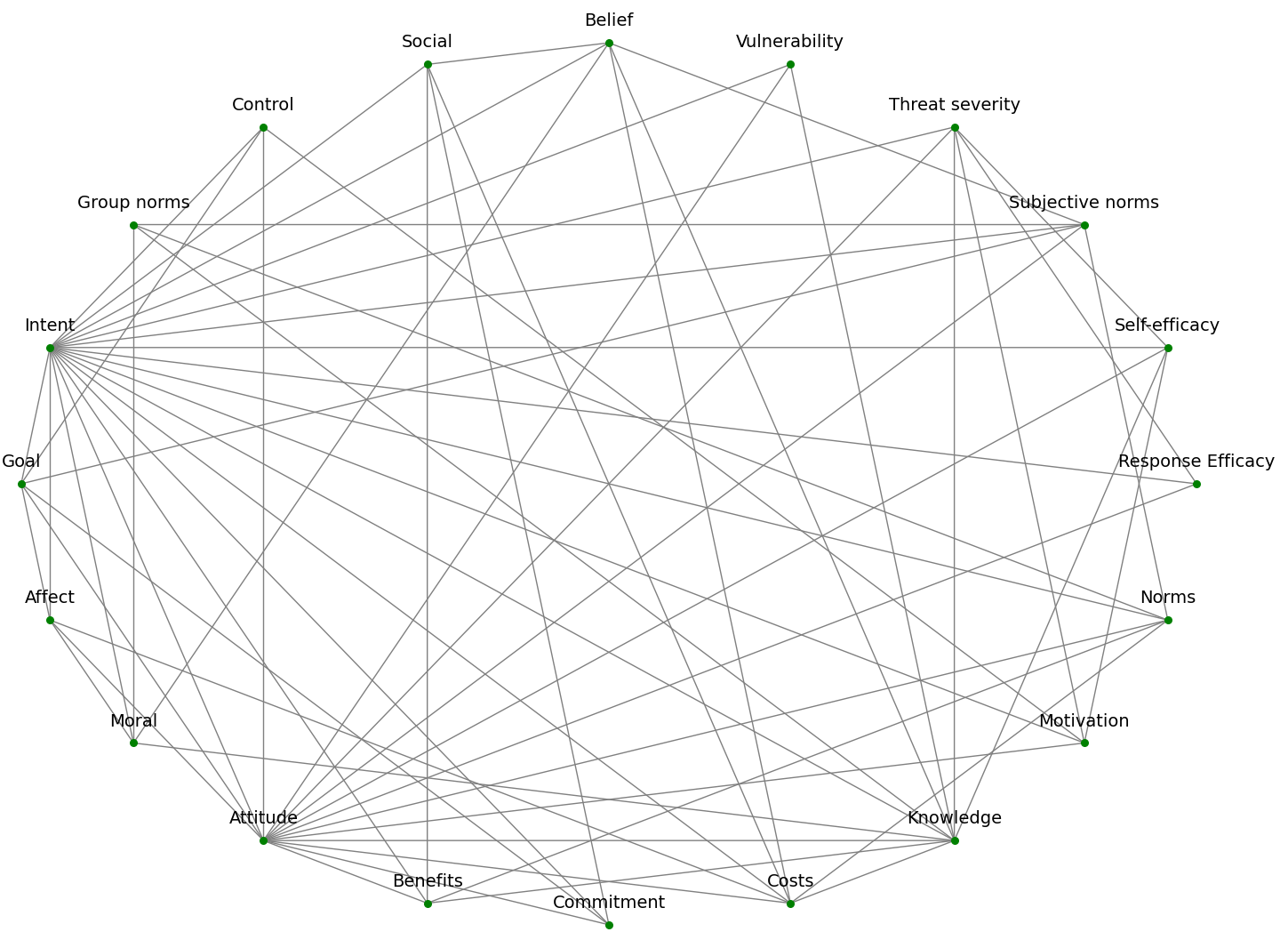}}
\caption{The Knowledge Graph of OllaBench}
\label{fig_kgraph}
\end{figure*}

\subsection{On Measuring Reasoning Ability}
In contemporary psychology, deductive and inductive reasoning are the main forms of reasoning. Deductive reasoning involves deriving a conclusion that is true provided the premises are accurate \cite{Wilhelm2005MeasuringAbility}. Conversely, inductive reasoning entails extrapolating beyond the given premises to generate conclusions of new semantic content\cite{Wilhelm2005MeasuringAbility}. Notably, people differ in their capacity to reason. While theoretically capable of rational thought, humans exhibit varying degrees of failure in practical reasoning. In principle, individuals accept reasoning as valid when they do not possess a conflicting mental representation \cite{Johnson-Laird1993ModelsRationality.,Stanovich1999WhoReasoning}.

The mental model theory, extensively utilized in understanding both deductive and inductive reasoning \cite{Johnson-Laird2015ModelsDeduction}, conceptualizes thinking as the manipulation of internal models \cite{Craik1967TheExplanation}. Entities are represented as tokens while relationships connect the tokens. Negations and implicit information are integrated within these models, shaping the reasoning process. The complexity of reasoning tasks is significantly influenced by the number of mental models congruent with the given premises \cite{Wilhelm2005MeasuringAbility}. Unlike the typical psychometric approach, which prioritizes test construction based on statistical criteria, a theory-driven approach would derive test indicators directly from a cognitive model of reasoning processes.

Additionally, Spearman's perspective on intelligence emphasizes the ability to identify relationships and patterns, a skill central to reasoning and reflected in its strong association with general intelligence metrics\cite{Wilhelm2005MeasuringAbility}. Also, Carroll's framework \cite{Carroll1993HumanStudies} identifies three reasoning dimensions: Sequential Reasoning, where conclusions are deduced from given premises or conditions; Induction, involving the detection and application of underlying rules or patterns in presented materials; and Quantitative Reasoning, which entails logical operations with numerical or mathematical concepts.

\subsection{On Evaluation of Large Language Models}
Language models, fundamentally designed to process and generate text, become highly versatile when trained on extensive datasets \cite{Bommasani2023HolisticModels}. This versatility underscores the necessity for comprehensive benchmarking to ensure their efficacy and transparency across various applications. Evaluation requires defining specific scenarios, a model, and metrics to assess performance. These evaluations span a broad spectrum of user-centric tasks, including question answering \cite{Kwiatkowski2019NaturalResearch, Kocisky2018TheChallenge, Choi2018QUAC:Context}, information retrieval \cite{Manning2009AnRetrieval}, summarizing, sentiment analysis, and toxicity detection.

While accuracy is crucial for system reliability \cite{Raji2022TheFunctionality}, it alone does not ensure AI models’ utility or desirability. The integration of machine learning models into larger systems demands not only accuracy but also the capability to express uncertainty—vital for error anticipation and management, particularly in high-stakes applications like decision-making processes. Calibration \cite{DeGroot1983TheForecasters}, or a model's ability to accurately estimate the probabilities of its predictions, is essential for deploying models in sensitive settings. This is quantitatively assessed through the expected calibration error (ECE), which gauges the alignment between predicted probabilities and actual outcomes \cite{Naeini2015ObtainingBinning, Guo2017OnNetworks}.

\section{OllaBench} \label{sec_ollabench}
\subsection{Theoretical Alignment}
Evaluation begins with setting the context and the overarching context was presented in figure \ref{fig_HIC_Framework}. The paper further narrow this context into the cognitive behavioral compliance and non-compliance of information security policies. With rising geopolitical tensions and state-sponsored mis/dis-information operations, divisive domestic political campaigns, persistent inflation, and growing lay-offs, the paper predicts cognitive behavioral compliance and non-compliance of information security policies is a timely focus for developing LLM-based applications in support of monitoring, modeling, and predicting the risk of insider threats. This initial scoping further involves identifying the main audience who are researchers and application developers being at the early stage of their LLM-based projects. They mainly need to pick the best LLM for successful project development and most efficient outcomes. Therefore, the chosen methodology is formative evaluation and the evaluative criteria for successful metrics only focus on relevance, alignment, effectiveness, and unintended effects. 

Knowledge selection is the next step. Heylighen \cite{Heylighen1997ObjectiveKnowledge} identified three principal classes of criteria for knowledge selection: objective, subjective, and inter-subjective. Objective criteria evaluate the "objectivity" or "reality" of knowledge. Subjective criteria concern the ease with which an evaluation target can assimilate knowledge. Inter-subjective criteria pertain to the ease of knowledge transmission and assimilation. Identifying the knowledge completeness is always difficult. The paper decided to employ the theory-driven method to formalize the knowledge base in the form of a knowledge graph which is a less formal form of ontology and is the right form for evaluation based on program theory \cite{Coryn2011A2009}. This decision also allows for the highest degree of objectivity to best support the goal of evaluating LLMs' subjective responses. The inter-subjective criteria is not currently in focus and will be explored in future projects.

Chosen evaluative metrics are exclusively quantitative-subjective. This decision follows the recommended practices \cite{PSU2013MetricsEvaluation} and was dictated by the selected scope and the underlying data. The metrics evaluate both deductive and inductive reasoning capabilities of the targeted LLMs. Notably, the paper assumes the LLMs have indexed the theories and related empirical evidences that belonged to the selected knowledge base. Based on the mental model theory \cite{Johnson-Laird2015ModelsDeduction}, the more difficult evaluative questions require the LLMs to construct more complex internal cognitive models for correct answers.

\subsection{The Mechanisms}
This section presents how theoretical alignment principles translate into actual implementation of OllaBench which has four main components of the knowledge graph, the scenario generator, the response generator, and the evaluator (figure \ref{fig_design}). Customization of the component parameters are possible via OllaBench's graphic user interface (GUI), changing parameters in json files, or modifying code variable values. The codes are publicly available on Github \footnote{https://github.com/Cybonto/OllaBench}. 

\subsubsection{The Knowledge Graph} The following theories informed the nodes and edges of the knowledge graph (a.k.a Cybonto-4-IC): Deterrence Theory \cite{Hu2011DoesEmployees}, 
Elaboration Likelihood Model \cite{Petty1986ThePersuasion}, 
Extended Parallel Processing Model \cite{Witte1992PuttingModel.}, 
Fear Appeal Theory \cite{Williams2012FearTheory}, 
General Deterrence Theory \cite{Herath2009ProtectionOrganisations}, 
IT Relatedness Theory \cite{Tanriverdi2005InformationFirms}, 
Neutralization Theory \cite{Sykes1957TechniquesDelinquency}, 
Organizational Support Theory \cite{Settoon1996SocialReciprocity.}, 
Protection Motivation Theory \cite{Rogers1997ProtectionTheory.}, 
Rational Choice Theory \cite{Elster1982Rationality}, 
Self-Determination Theory \cite{Deci2012Self-determinationTheory.}, 
Self-Efficacy Theory \cite{Bandura1997Self-efficacy:Control.}, 
Situation Action Theory \cite{Wikstrom2009ViolenceAction}, 
Social Bond Theory \cite{Hirschi2017CausesDelinquency}, 
Social Cognitive Learning Theory \cite{Bandura2001SocialPerspective}, 
Social Exchange Theory \cite{Blau2017ExchangeLife}, 
Social Norms Theory \cite{berkowitz2005overview}, 
Technology Threat Avoidance Theory \cite{Liang2009AvoidancePerspective}, 
Theory of Acceptance Model \cite{Venkatesh2003UserView}, 
Theory of Computer Crime Opportunity Structure \cite{Willison2006OpportunitiesPerspective}, 
Theory of Emotion Process \cite{Frijda1989RelationsReadiness.}, 
Theory of Interpersonal Behavior \cite{Triandis1979ValuesBehavior.}, 
Theory of Planned Behavior \cite{Ajzen1991TheBehavior}, and 
Theory of Reasoned Action \cite{Fishbein1980AImplications.}.

The nodes are the cognitive behavioral constructs while the edges are the relationships among the nodes, all of which were proposed by the theories. The paper further looked up 38 peer-reviewed papers to gather empirical supports for the edges. Empirical evidence is usually gathered using statistical approaches involving factor analysis and multiple regression analysis such as Structural Equation Modeling \cite{Ullman2012StructuralModeling}. Further details can be found at the project's online Cybonto-4-InterdendentCybersecurity page \cite{Nguyen2024CYBONTO-1.0/Cybonto-4-InterdependentCybersecurity.mdCybonto/CYBONTO-1.0}.

\subsubsection{The Scenario Generator} 
Each node has two set of seed description - one for information security policy compliance and another one for non-compliance. A set of seed description for each node (a cognitive behavioral construct) contains at least two sentences describing the cognitive behavioral details specific to the node. These sentences came from the validated and evaluated survey instruments used by the peer-reviewed papers. Each sentence was mutated into a library of similar sentences using an LLM - the paper used GPT-4. These sentences form a verbal description library for each node. Below is one example of how seeds were mutated. Seed and verbal description library datasets are also available on the project's Github repository.

\begin{tcolorbox}
\textbf{Sample mutation of seeds:}\\
Construct: Belief \\
Compliant seed: The person believes that rules are made to be followed.\\
Compliant mutation: The individual believes in strict adherence to rules and regulations.\\
Non-compliant seed: The person believes that it’s ok to get around a policy if the person can get away with it.\\
Non-compliant mutation: The individual sees loopholes in rules as acceptable if they go unnoticed.
\end{tcolorbox} 

A cognitive behavioral profile generation begins with selecting a random path of length $n$ from the knowledge graph. The paper recommends $n=4$ which yields a list of 5 nodes. Nodes are then assigned properties of compliant or non-compliant to information security policies. The maximum and minimum numbers of non-compliant nodes in a cognitive behavioral profile are 5 and 0 respectively. From the list of selected nodes, node properties, and corresponding verbal description libraries, the verbalized cognitive behavioral profiles are made. This modular approach allows users to customize scenarios for their evaluation purposes. Each scenario contains two cognitive profiles of two hypothetical employees. The standard benchmark dataset generated by this paper has 10,000 unique scenarios.

\subsubsection{The Response Generator} For each unique scenario, the response generator asks the targeted LLMs the same set of multiple (4) choice questions as follows:
\begin{itemize}
    \item Which of the following options best reflects [Person A]'s or [Person B]'s cognitive behavioral constructs?
    \item Who is [LESS/MORE] compliant with information security policies?
    \item Will information security non-compliance risk level [increase/decrease] if these employees work closely in the same team?
    \item To [increase information security compliance/reduce information security non-compliance], which cognitive behavioral factor should be targeted for strengthening?
\end{itemize}

Each choice begins with "(option [a/b/c/d])" and LLMs were instructed to begin their answers with "(option [a/b/c/d])". The instruction does not limit LLMs' responses which means LLMs are free to elaborate on their answer choices as they wish.

\begin{tcolorbox}
\textbf{A sample cognitive behavioral profile:}\\
Here are the intelligence about Olivia Hernandez with comments from trusted experts and/or Olivia Hernandez's recorded statement(s):
\begin{itemize}
    \item It comes easily to the individual to utilize cybersecurity protection software.
    \item The individual is knowledgeable about the breaches and repercussions of the company's protocols.
    \item The individual is capable of easily utilizing security software for online protection.
    \item The individual views following the institution's policies regarding information security as compulsory.
    \item My disregard for information security policies could benefit me.
\end{itemize}
\end{tcolorbox} 

\subsubsection{The Evaluator} For each correct response, the evaluator gives a score of 1 and 0 vice versa. This grading involves comparing LLMs' responses with reference answers which can be reliably constructed as follows.

For question 1 (a.k.a the Which-Cog-Path question), the reference answer is the list of nodes based on the selected path used for the cognitive behavioral profile of either person A or B. The choices includes some permutations of the correct answer. To answer this question correctly, the LLMs must be able to map each description sentence back to the right construct which is possible if the LLMs are aware of the relevant cognitive behavioral instruments used by numerous peer-reviewed papers.

For question 2 (a.k.a the Who-is-Who question), person A is more compliant than person B if A has more nodes of compliant attribute than B. The LLMs must be able to count the numbers of compliant and non-compliant attributes of the nodes (cognitive behavioral constructs) in each profile to compare them. There is appropriate option for the LLMs to choose if A and B share the same numbers of node attributes. 

For question 3 (a.k.a the Team-Risk question), there are 3 possibilities. First, if both individuals A and B have no nodes with non-compliant attributes, there will be no change in the team's compliance risk profile. Second, if A and B possess non-overlapping nodes with non-compliant attributes, this results in a higher aggregate number of non-compliant nodes, thereby increasing the non-compliant risk profile should they collaborate in a team. Third, if A and B share the same non-compliant nodes, the impact on the non-compliant risk profile may vary. The main rationale for these dynamics is that team members tend to influence and adopt each other’s behaviors, including potentially harmful cognitive behavioral patterns. This assertion is supported by various theoretical frameworks including the Social Cognitive Learning Theory \cite{Bandura2001SocialPerspective}, Social Norms Theory \cite{berkowitz2005overview}, Theory of Interpersonal Behavior \cite{Triandis1979ValuesBehavior.}, and Theory of Computer Crime Opportunity Structure \cite{Willison2006OpportunitiesPerspective}.

For question 4 (a.k.a the Target-Factor question), if A and B shares one non-compliant node then that cognitive behavioral construct (node) is the correct answer. For other possibilities, the target construct for strengthening should be the node (construct) with the highest page-rank among the combined non-compliant nodes or the combined compliant nodes (if there is no non-compliant node). In terms of difficulty, question 4 is the hardest since it requires the LLMs to think beyond the information presented by the scenario. The LLM that can construct the most robust networks around the involved nodes will perform the best in answering this question. 

\subsection{The Metrics}
The paper denotes $A_{WCP}, A_{WHO}, A_{TR}, A_{TF}$ as the accuracy values when a model responses to Which-Cog-Path questions, Who-is-Who questions, Team-Risk questions, and Target-Factor questions respectively. Such values can also be called as "Categorical Accuracy" or $A_{cat}$. In a category $cat$, the number of questions is $n_{cat}$ and the number of correct responses is $r_{cat}$. The formula for calculating categorical accuracy is
\begin{equation} 
    \NORMAL{ A_{cat}=\frac{r_{cat}}{n_{cat}} }
\end{equation} 
The overall Accuracy score is the average of the categorical accuracy values.

The paper denotes $W_x$ for LLM x's wastefulness value. $n_{missed}$ is the total number of the model's wrong answer. $T_i$ is the tokens used to generate a wrong answer $i$. The wastefulness of a model x is
\begin{equation} 
    \NORMAL{ W_x=\frac{\sum_{i=0}^{n_{missed}-1}T_i}{n_{missed}} }
\end{equation} 

\begin{figure}[ht]
\centerline{\includegraphics[width=220pt,height=\textheight,keepaspectratio]{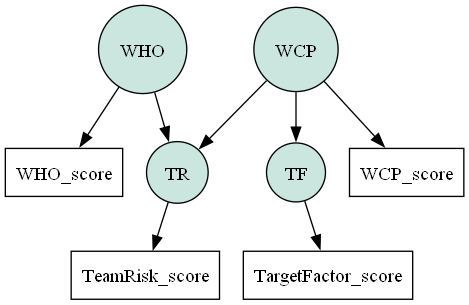}}
\caption{Consistency Model}
\label{fig_consistency}
\end{figure}

Finally, the consistency metrics measures the possibility of whether the LLMs' responses are consistent with a knowledge model. To achieve this, the paper employs Structural Equation Modeling \cite{Ullman2012StructuralModeling}, a comprehensive statistical approach used to assess complex relationships among variables. Initially, the paper propose a model (figure \ref{fig_consistency}) which is based on the paper's subjective knowledge of the domain.

\begin{figure*}[ht]
\centerline{\includegraphics[width=450pt,height=\textheight,keepaspectratio]{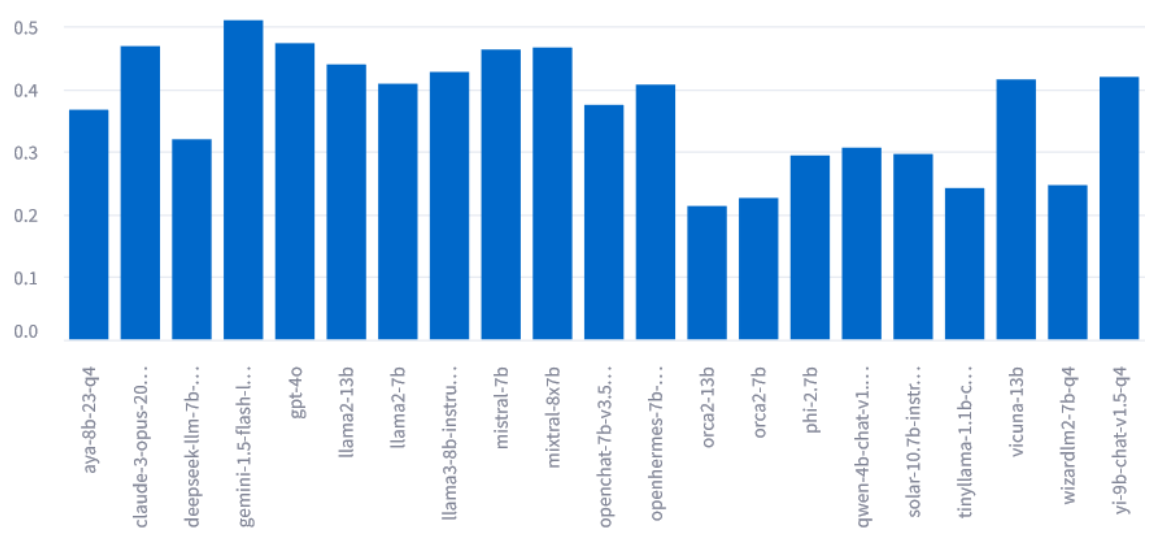}}
\caption{Overall Accuracy}
\label{fig_accuracy}
\end{figure*}

\begin{figure*}[ht]
\centerline{\includegraphics[width=450pt,height=\textheight,keepaspectratio]{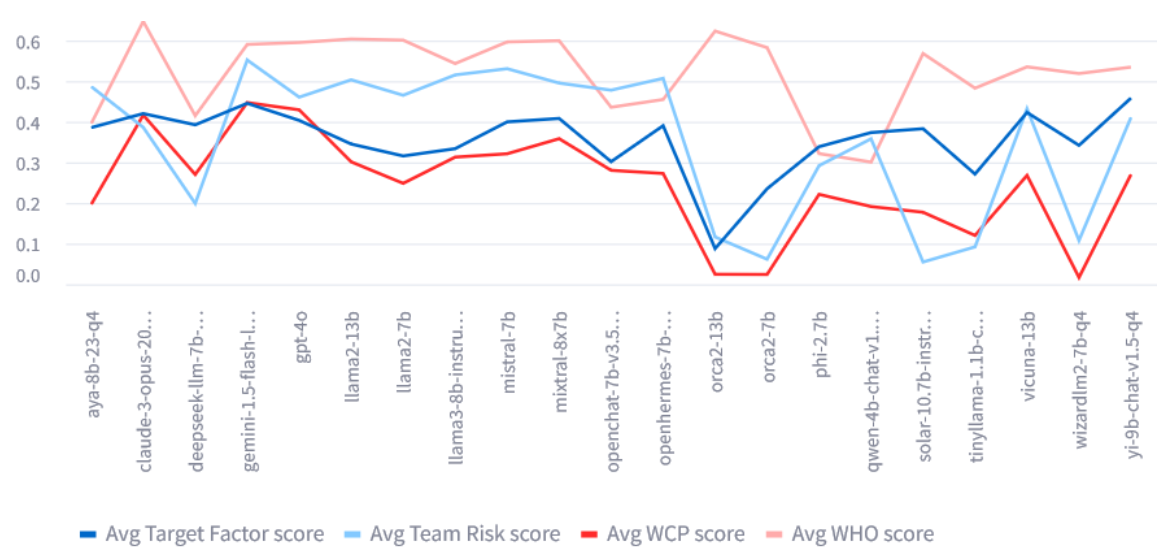}}
\caption{Categorical Accuracy}
\label{fig_cat_ccuracy}
\end{figure*}

\begin{figure*}[ht]
\centerline{\includegraphics[width=450pt,height=\textheight,keepaspectratio]{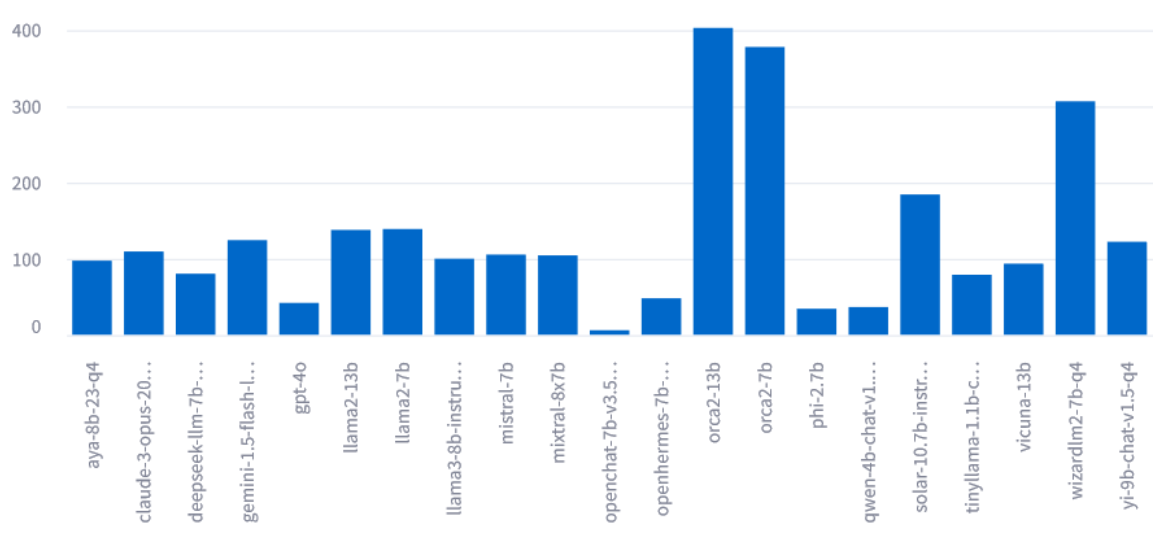}}
\caption{Wastefulness}
\label{fig_wastefulness}
\end{figure*}

In Figure \ref{fig_consistency}, "WHO," "WCP," "TR," and "TF" represent scores from the LLMs' answers to the Who-is-Who questions, the Which-Cognitive-Path questions, the Team-Risk questions, and the Target-Factor questions, respectively. A directed arrow from node A to node B indicates that the knowledge used in answering the node A question contributes to the knowledge used in answering the node B question. The model is most consistent when it answers both node A and B questions correctly or incorrectly. Conversely, the model is least consistent when it answers the node A question incorrectly and the node B question correctly. The, Semopy (a Python package for performing Structural Equation Modeling) was used with Wishart Loglikelihood for objevtive function (Semopy default setting) on LLMs' response scores.

The Structural Equation Modeling process yields main metrics of Estimate, Standard Error (Std. Err), z-value, and p-value. Estimate indicates the strength and direction of the relationship between variables - larger absolute values mean stronger relationships. Standard error measures the precision of the estimate - smaller values mean more precise estimates. Larger absolute z-values suggest stronger relationships. Finally, the p-value indicates the statistical significance of the relationship and should be small (typically less than 0.05).

\section{Evaluating LLMs Results} \label{sec_results}
\begin{tcolorbox}
\textbf{Key findings:}\\
$\bullet$ Commercial LLMs have the highest overall accuracy scores as expected, although the scores should be higher.\\
$\bullet$ Smaller and low-resolution open-weight LLMs are not far from the leading commercial LLMs.\\
$\bullet$ There are large margins between the most efficient LLMs and the most wasteful LLMs in terms of tokens spent in wrong answers.\\
$\bullet$ The most accurate LLMs are most consistent in answering questions.\\
\end{tcolorbox}
OllaBench was used to evaluate the following LLM models:
\begin{enumerate}
    \item aya-8b-23-q4 (released by Cohere)
    \item claude-3-opus-20240229 (released by Anthropic)
    \item deepseek-llm-7b-chat-q4 (released by Deepseek )
    \item gemini-1.5-flash-latest (released by Google)
    \item gpt-4o (released by OpenAI)
    \item llama2-13b (released by Meta)
    \item llama2-7b (released by Meta)
    \item llama3-8b-instruct-q4 (released by Meta)
    \item mistral-7b (released by Mistral AI)
    \item mixtral-8x7b (released by Mistral AI)
    \item openchat-7b-v3.5-q4 (released by OpenChat)
    \item openhermes-7b-mistral-v2.5-q4 (released by Teknium)
    \item orca2-13b (released by Microsoft Research)
    \item orca2-7b (released by Microsoft Research)
    \item phi-2.7b (released by Microsoft Research)
    \item qwen-4b-chat-v1.5-q4 (released by Alibaba Cloud)
    \item solar-10.7b-instruct-v1-q4 (released by Upstage)
    \item tinyllama-1.1b-chat-v0.6-q4 (released by Peiyuan Zhang et al.)
    \item vicuna-13b (released by LMSYS org)
    \item wizardlm2-7b-q4 (released by Microsoft AI)
    \item yi-9b-chat-v1.5-q4 (released by 01.AI)
\end{enumerate}
All open-weight LLMs are 4-bit quantized.

\begin{table*}[ht]
\centering
\caption{Structural Equation Modeling Results - Top 5}
\label{tab:sem}
\resizebox{\textwidth}{!}{%
\begin{tabular}{@{}llrrrr@{}}
\toprule
 & \textbf{Model} & \textbf{Estimate} & \textbf{Std. Err} & \textbf{z-value} & \textbf{p-value} \\ \midrule
\multirow{5}{*}{\textbf{Team Risk $\sim$Which Cognitive Path}} & mixtral-8x7b & -0.952 & 0.0006 & -1,546.09 & 0 \\
 & openchat-7b-v3.5-q4 & -0.9541 & 0 & -19,361.53 & 0 \\
 & orca2-13b & -0.9398 & 0 & -36,803.24 & 0 \\
 & yi-9b-chat-v1.5-q4 & 1.3453 & 0 & 1,345,339.47 & 0 \\
 & wizardlm2-7b-q4 & 1.524 & 0.6844 & 2.2267 & 0.026 \\ \midrule
\multirow{5}{*}{\textbf{Team Risk $\sim$Who is Who}} & gemini-1.5-flash-latest & -0.1225 & 0.0136 & -9.0078 & 0 \\
 & gpt-4o & 0.172 & 0.0094 & 18.2274 & 0 \\
 & llama3-8b-instruct-q4 & 0.0848 & 0.0082 & 10.3026 & 0 \\
 & llama2-7b & 0.0846 & 0.0183 & 4.6247 & 0 \\
 & tinyllama-1.1b-chat-v0.6-q4 & 0.0276 & 0.0123 & 2.2454 & 0.0247 \\ \midrule
\multirow{5}{*}{\textbf{Target Factor $\sim$Which Cognitive Path}} & mixtral-8x7b & 0.8662 & 0.9574 & 0.9047 & 0.3656 \\
 & openchat-7b-v3.5-q4 & 0.8647 & 0.9936 & 0.8702 & 0.3842 \\
 & openhermes-7b-mistral-v2.5-q4 & 0.8648 & 1.0662 & 0.8111 & 0.4173 \\
 & mistral-7b & 0.1431 & 0.1831 & 0.7819 & 0.4343 \\
 & orca2-7b & 0.5436 & 1.5621 & 0.348 & 0.7278 \\ \bottomrule
\end{tabular}%
}
\end{table*}

\textbf{The top 5 LLMs in answering Which-Cognitive-Path questions are:}
\begin{enumerate}
    \item gemini-1.5-flash-latest (0.4485)
    \item gpt-4o (0.4306)
    \item claude-3-opus-20240229 (0.4177)
    \item mixtral-8x7b (0.3593)
    \item mistral-7b (0.3221)
\end{enumerate}

\textbf{The top 5 LLMs in answering Who-is-Who questions are:}
\begin{enumerate}
    \item claude-3-opus-20240229 (0.649)
    \item orca2-13b (0.6252)
    \item llama2-13b (0.6054)
    \item llama2-7b (0.6027)
    \item mixtral-8x7b (0.6013)
\end{enumerate}

\textbf{The top 5 LLMs in answering Team-Risk questions are:}
\begin{enumerate}
    \item gemini-1.5-flash-latest (0.5536)
    \item mistral-7b (0.5322)
    \item llama3-8b-instruct-q4 (0.5168)
    \item openhermes-7b-mistral-v2.5-q4 (0.5083)
    \item llama2-13b (0.5047)
\end{enumerate}

\textbf{The top 5 LLMs in answering Target-Factor questions are:}
\begin{enumerate}
    \item yi-9b-chat-v1.5-q4 (0.4597)
    \item gemini-1.5-flash-latest (0.4467)
    \item vicuna-13b (0.4241)
    \item claude-3-opus-20240229 (0.4213)
    \item mixtral-8x7b (0.4093)
\end{enumerate}

\textbf{The top 5 LLMs in overall accuracy are:}
\begin{enumerate}
    \item gemini-1.5-flash-latest (0.5102)
    \item gpt-4o (0.4735)
    \item claude-3-opus-20240229 (0.4688)
    \item mixtral-8x7b (0.4667)
    \item mistral-7b (0.4634)
\end{enumerate}

\textbf{The top 5 LLMs in efficiency (lowest in wastefulness) are:}
\begin{enumerate}
    \item openchat-7b-v3.5-q4 (6.427 tokens)
    \item phi-2.7b (34.5295 tokens)
    \item qwen-4b-chat-v1.5-q4 (36.7592 tokens)
    \item gpt-4o (42.2121 tokens)
    \item openhermes-7b-mistral-v2.5-q4 (48.2287 tokens)
\end{enumerate}

\textbf{The top 5 LLMs in wastefulness are:}
\begin{enumerate}
    \item orca2-13b (402.038 tokens)
    \item orca2-7b (377.1546 tokens)
    \item wizardlm2-7b-q4 (305.9823 tokens)
    \item solar-10.7b-instruct-v1-q4 (184.0692 tokens)
    \item llama2-7b (138.885 tokens)
\end{enumerate}

Figure \ref{fig_cat_ccuracy} show the LLMs performance in categorical accuracy.
Figure \ref{fig_accuracy} shows the overall LLMs accuracy in answering questions.
Figure \ref{fig_wastefulness} shows the LLMs wastefulness (the lower the better).
Full benchmark results are available at the project's Github page \footnote{https://github.com/Cybonto/OllaBench/tree/main/BenchmarkResults}


\section{Conclusion}
Large Language Models (LLMs) can be highly useful in building realistic agents for Agent-Based Modeling (ABM).  ABM agents can represent subsystems and their entities, governed by flexible learnt rules that facilitate the simulation of complex relationships and interactions \cite{An2020Editorial:Models, Silva2020SimulatingReview}. This bottom-up approach allows the micro-level agent behaviors to collectively generate emergent macro-level structures. This unique capability may help with addressing Interdependent Cybersecurity problems that involve complex networks of humans, infrastructures, technologies, finance, and so on.

Evaluation of LLMs is required by law especially when the LLMs' decisions may affect humans. Evaluation of LLMs is also essential for LLM-based application development and commercial deployment. We notice that most recent LLM evaluation works for cybersecurity does not involve human factor while LLM reasoning evaluation works do not involve both cybersecurity and human factor. This phenomenon is consistent with the fact that research and applications focusing on human-centered IC remains limited \cite{Kianpour2021SystematicallySurvey} while contemporary LLM evaluation frameworks, such as Stanford's HELM \cite{Bommasani2023HolisticModels}, fail to assess the cognitive computing capabilities of LLMs.

Therefore, OllaBench was born to help both researchers and application developers conveniently evaluate their LLM models within the context of cybersecurity compliance or non-compliance behaviors. Specifically, OllaBench can rank LLMs based on their accuracy, wastefulness, and consistency in answering scenario-based information security compliance/non-compliance questions. After applying OllaBench in evaluating 21 LLMs including major commercial LLMS, the key findings are:\\
$\bullet$ Commercial LLMs have the highest overall accuracy scores as expected, although the scores should be higher.\\
$\bullet$ Smaller and low-resolution open-weight LLMs are not far from the leading commercial LLMs.\\
$\bullet$ There are large margins between the most efficient LLMs and the most wasteful LLMs in term of tokens spent in wrong answers.\\
$\bullet$ The most accurate LLMs are most consistent in answering questions.

OllaBench results are reliable because OllaBench was rigorously designed based on a solid foundation of 24 cognitive behavioral theories and empirical evidence from 38 peer-reviewed papers. OllaBench is also robust as it supports evaluation of all open-weight models and models from major platforms such as OpenAI, Anthropic, and Google. OllaBench is user-friendly with supported command-line interface and graphical user interface. Early stage solution developers may leverage OllaBench to identify the best LLM for their human-centric IC application development. Researchers may leverage OllaBench to measure their fine-tuned LLMs' performances. Customizing OllaBench for other human-centric purposes is convenient with carefully documented source codes publicly available on Github.

\bibliographystyle{IEEEtran}
\bibliography{IEEEabrv,references.bib}

\end{document}